\font\title=cmssdc10 scaled 1200
\magnification=\magstep1
\baselineskip 22 true pt \parskip=0pt plus3pt
\hsize 5.5 true in \hoffset .125 true in
\vsize 8.5 true in \voffset .1 true in

 \def\Schrod{Schr\"{o}dinger }
\def\pmb#1{\setbox0=\hbox{#1}
 \kern-.025em\copy0\kern-\wd0
\kern.05em\copy0\kern-\wd0
 \kern-.025em\raise.0433em\box0 }

\def\ket#1{|#1\rangle}
\def\doubleket#1{||#1\rangle\rangle}
\def\bra#1{\langle #1|}

\def\kvac{\ket{0}}
\def\bvac{\bra{0}}
\def\bvacp{\bra{0^\prime}}
\def\eq#1{\eqno{(#1)}}

\centerline{{\bf {\title{How a Quantum Theory
 Based on Generalized Coherent States }}}}
\centerline{{\bf {\title{ Resolves the EPR and
Measurement Problems}}}}
\vskip.1in
\centerline{{\bf Daniel I. Fivel}}
\vskip.05in
\centerline{{\bf  Department of Physics}}
\centerline{{\bf University of Maryland, College Park,
MD 20742}}
\vskip.05in
\centerline{{\bf April 25, 2001}}
\vskip.2in
\centerline{{\bf Abstract}}
\vskip.2in
It is shown that the quantum theory can be formulated on
homogeneous spaces of generalized coherent states in a
manner that accounts for interference,
entanglement, and the linearity of dynamics
{\it without using the superposition principle}. The
coherent state labels, which are essentially
instructions for preparing states,  make it unnecessary
to identify properties with projectors in Hilbert space.
 This eliminates the so called
``eigenvalue-eigenstate" link, and the theory thereby
escapes the measurement problem. What the theory allows
us to predict is the distribution in the outcomes of
tests of {\it relations} between coherent states. It is
shown that quantum non-determinism can be attributed
to a hidden variable (noise) in the space of relations
without violating the no-go theorems (e.g.
Kochen-Specker). It is shown that the coherent state
vacuum is distorted when entangled states are generated.
The non-locality of the vacuum permits this distortion
to be felt everywhere without the transmission of a
signal and thereby accounts for EPR correlations in a
manifestly covariant way.
\vskip.2in
\vfill
\eject
  If a theorist wishes to instruct an
experimentalist to prepare a specific quantum state, it
wil be of little use to the experimentalist if he is
given a list of complex numbers that are to be its
components in some Hilbert space basis.
 What the experimentalist  requires is a {\it recipe}
for preparing the state, e.g.\ to turn a Stern-Gerlach
magnet in a prescribed way or to turn on
 a prescribed laser-generating current. In general the
recipes specify a transformation 
$g$ to be applied to a reference state, and the
transformations form a group ${\cal G}$ that is
characteristic of the system. For experiments with
Stern-Gerlach magnets ${\cal G}$ is the rotation group 
$SO_3$, and for laser states it is
the Weyl-Heisenberg (WH) group expressed by exponentials
of bose operators in which the current driving the laser
appears.
\vskip.05in
States described by such recipes are called
{\it generalized coherent states}. The Weyl-Heisenberg
states were first introduced into optics by
Glauber$^{\bf 1}$, and the concept was subsequently
generalized by Perelomov$^{\bf 2}$ who, with his
co-workers, discovered most of what we know about them.
\vskip.05in 
If generalized coherent states are all that 
experimentalists ever prepare, one is motivated to apply
Occam's razor to the theory, so that they alone appear in
it. But although the coherent states are complete (in
fact overcomplete), linear combinations of coherent
states are not in general coherent states. Hence, if we
are to permit the coherent states themselves and no
others, we shall have to abandon the superposition
principle. 
\vskip.05in
The superposition principle is a pillar of orthodox
quantum theory and is also the source of the measurement
problem. I will show that the phenomena which are taken
to justify the superposition principle (interference,
entanglement, linear dynamics) can be correctly accounted
for in a model, which I will call the coherent state
(CS) model,  which never requires us to  introduce
superpositions of coherent states. We  then will not be
required to identify properties with lattices of
subspaces in Hilbert space, and the measurement problem
will literally go away.  
\vskip.05in  
The construction of the CS model begins with the
hypothesis that for each  system there is a
characteristic group 
${\cal G}$ which will be called the
``coherence group" of the system. No restriction is
placed on
${\cal G}$ other than that it be locally compact, a
restriction quite in keeping with experimental
implementability. By assuming it we shall have the
benefit of an invariant (Haar) measure $d\mu$ by which we
can integrate over the group when we have to.
\vskip.05in Let
$g \in {\cal G} \to U(g)$ be an irreducible, unitary
representation on a Hilbert space
${\cal H}$. Let
$\kvac$ and $\bvacp$ be reference states for systems and
detectors respectively. Let
${\cal G}_o $ and ${\cal G}_o^\prime$ be the stability
subgroups for the system and detector reference states,
i.e.\ the subgroups that leave them invariant.
\vskip.05in
 A set
${\cal F}$ of generalized coherent system states and a
set ${\cal F}^\prime$ of generalized coherent detector
states are defined by
$$\ket{g} = U(g)\kvac,\quad \bra{g} = \bvacp
U^\dagger(g),\eq{1}$$ in which we select one element $g$
from each left coset $g{\cal G}_o$ for ${\cal F}$ and
one element
$g$ from each right coset ${\cal G}_o^\prime g$ for
${\cal F}^\prime$. Thus ${\cal F} = {\cal G}/{\cal G}_o$
and
${\cal F}^\prime = {\cal G}o^\prime\backslash {\cal G}$
are homogeneous spaces. Note that
$\bra{g}$ is not the dual of
$\ket{g}$ unless it happens that the two reference
states are duals of one another.©Ä
\vskip.05in The dramatic effect of restricting the
allowed states to generalized coherent states  is that
{\it the group structure of
${\cal G}$ is imparted to quantum mechanical amplitudes}
in the following way:
$$\bra{g_1}g_2\rangle = \bra{0^\prime}
U^\dagger(g_1)U(g_2) \kvac = 
\bra{0^\prime} U^{-1}(g_1)U(g_2) \kvac =
\bvacp U(g_1^{-1})U(g_2) \kvac = $$ $$
\bra{0^\prime} U(g_1^{-1}g_2) \kvac =
\bra{0^\prime} U(g) \kvac 
 \equiv f(g),\; g = g_1^{-1}g_2.\eq{2}$$ All predictions
will be obtained from the amplitude $f(g)$ which is a
function on
${\cal G}$ that is constant on each  double coset 
${\cal G}_0^\prime \backslash {\cal G}/{\cal G}_0$. The
double cosets partition ${\cal G}$ just as left and
right cosets do$^{\bf 3}$, and so we may treat the set
${\cal S}$ of double cosets as the {\it set of 
possible relations} between system and detector states.
Each pair
$g_1,g_2$ for which $g_1^{-1}g_2 = g $ belongs to the
same double coset may be regarded as different {\it
manifestations} of the relation
$g$.\vskip.1in We see then that {\it when restricted to
coherent states the quantum theory can be regarded as a
theory of relations rather than states, and these
relations have a group theoretic structure.} In
particular
 the fundamental rule, that
$|\bra{g_2}g_1\rangle|^2$ is the  probability for a
system in state
$\ket{g_1}$ to pass a detector in state
$\bra{g_2}$, now becomes a rule for computing the
probability that the relation $g$ ``holds", and we write:
$$p(g) = |f(g)|^2.\eq{3}$$

This leads us to the next benefit of our restriction to
coherent states, for we observe that the probability
function has a suggestive geometric structure. Since
$\bra{g_1}g_2\rangle$ is a scalar product, the function
$$s(g_1,g_2) \equiv \sqrt{1 -
|\bra{g_1}g_2\rangle|^2}\eq{4}$$ is a metric on
${\cal G}.$ This distance between $g_1$ and
$g_2$ is the same as the distance of $g = g_1^{-1}g_2$
to the identity of the group. Thus we can interpret 
$$s(g) \equiv s(g,e) = \sqrt{1 - |f(g)|^2}\eq{5}$$ as
the ``size" of the relation, and its square as a ``cross
section" for the relation. Thus one may think of
experiments that test the relation in the way one thinks
of scattering experiments. We ``throw" a random relation
$h$ at $g$ and say that $g$ holds if it is smaller than
$h$, i.e.\ if $s(g) < s(h)$. Thus the probability that
$g$ holds is the probability that we have thrown an $h$
which is bigger than $g$. If we throw $h$'s with a
distribution such that the probability of having $s(h) <
r$ is the cross-section of a disk of radius $r$, then
the probability that
$g$ holds is just the value $1 - (s(g))^2 = |f(g)|^2$
required by quantum mechanics.
\vskip.05in
 We define a random variable with values in
${\cal S}$ that is  distributed in this way as the
``relational hidden variable" for ${\cal G}$ coherent
states. It should be noted that there is no conflict
with no-go theorems because relations are intrinsically
non-local. The existence of this kind of hidden variable
 was observed long ago by Bell$^{\bf 4}$ for spin-1/2
systems. What we see here is that there is a natural
generalization of the idea to all coherent state systems.
\vskip.05in 
 We pause here to examine the
two most important examples for experimental
applications .
\vskip.05in 
\noindent {\it Example 1} ---  Detection of spin-1/2
particles with Stern-Gerlach magnets:
${\cal G} = SO_3$. 
$U$ is the two dimensional representation. Taking any
state as the reference state, the stability subgroup is
the
 rotation about that direction. Thus ${\cal F} =
SO_3/SO_2 = SU_2/U_1$ which is the 2-sphere. Thus states
are labeled by their coordinates ($\theta,\phi$), and if
the reference state for both system and detector is the
north pole we find that
$s(\theta,\phi) = \sin(\theta/2)$ (the chord metric).
Since the fraction of the area of a sphere lying within
polar latitude
$\theta$ of the pole is $\sin^2(\theta/2)$,  we see that
the relational hidden variable distribution is (happily)
uniform over the area of the sphere.
\vskip.05in Note that in this example every ray
corresponds to a coherent state. This will be the case
whenever
${\cal G}$ is the full unitary group
$U_N$ in
$N$-dimensions because any unit vector can be obtained
from any other unit vector by a unitary transformation.
One can still not say that the set of coherent states is
{\it linearly} closed because  of the neccessity of
normalizing linear combinations to make them state vectors.
\vskip.05in The  groups
$U_N$ for arbitrarily large but finite $N$ are compact,
but for
  $N \to \infty$ the group is not even locally compact.
Our restriction to locally compact groups thus
recognizes the increasing difficulty of experimentally
implementing the full unitary group as its dimension
increases. 
\vskip.05in
 \noindent {\it Example 2} --- Detection of the complex
amplitude
$\lambda$ of a single mode laser beam by the
photocurrent it produces: The coherent states are the 
Glauber states.
${\cal G}$ is the Weyl-Heisenberg group (WH) in which
the group elements have the composition law
$$	g = e^{i\theta}U(\lambda), \; \lambda \in {\cal
C},\qquad U(\lambda_1)U(\lambda_2) =
e^{iIm(\lambda_2^*\cdot\lambda_1)}U(\lambda_1 +
\lambda_2).\eq{6}$$  In  coherent superpositions it is
the
$\lambda$'s that  are added, so the important lesson
learned from (6) is that {\it coherent superpositions
are described by group multiplication 
 not state vector addition.} 
\vskip.05in By the Stone-von Neumann Theorem there is
only one irreducible representation up to equivalence,
the so-called Fock representation. In this
representation $U$ is expressed in terms of bose
operators
$a,a^\dagger$ with
$[a,a^\dagger] = 1$: 
$$U(\lambda) = e^{\lambda a^\dagger -
\lambda^* a}.\eq{7}$$ The reference state
$\kvac$, called the Fock vacuum, is annihilated by
$a$, and its stability subgroup is $U_1$ (phase
multiplication). The space ${\cal F} = WH/U_1$ is the
single mode phase space, i.e.\ the homogeneous space is
the complex plane. Thus the coherent system states are
$$\ket{\lambda} = U(\lambda)\kvac, \; \lambda
\in {\cal C}
\Longrightarrow \bra{0}U(\lambda)\ket{0} =
e^{-|\lambda|^2/2}.\eq{8}$$
In the coherent detector states $\bra{\lambda}$, the
parameter $\lambda$ is the detected photocurrent. 
\vskip.05in
 The
invariant measure on WH is the area measure in the
complex plane, so the measure on the group for which
$\lambda$ is in the annulus of width
$d|\lambda|$ at
$|\lambda|$ is 
$dA = \pi d|\lambda|^2.$ If the distribution  of a trial
value
$\mu $ with respect to the measure is
$\pi^{-1}e^{-|\mu|^2}d|A$, then the probability for
$|\mu| > |\lambda|$ is $e^{-|\lambda|^2}$ which is the
squared modulus of (8). It follows that 
{\it the relational hidden variable for WH groups is
Maxwellian }
 i.e.\ it represents ``thermal" noise in the space of
relations.
\vskip.05in
 The WH group is non-compact but is locally compact. The
direct product $WH_N$ of an arbitrarily large but finite
number $N$ of WH groups (which describes an
$N$-mode laser) is also locally compact. We may
represent it by simply replacing the complex number
$\lambda$ with an
$N$ component complex vector and understand
$\lambda\cdot a$ to be $\lambda_1 a_1 +
\cdots +
\lambda_N a_N$, in which the $a_j$'s are commuting bose
operators, and the composition law (6) holds with
$\lambda_2^*\cdot\lambda_1$
 being the complex scalar product of the vectors
$\lambda_1$ and
$\lambda_2$.
\vskip.05in  Up to this point we have said nothing about
dynamics, and it is the linearity of quantum dynamics
that has been the principle justification for the
superposition principle. In order to keep dynamics
linear when we restrict the states to be coherent, we
can only allow those linear transformations that 
preserve the group structure (automorphisms) and are
smooth (homeomorphisms) to preserve the topological
structure of the homogeneous spaces
${\cal F}$ and
${\cal F}^\prime$. Such transformations will be 
implemented in
${\cal H}$ by unitary transformations
$A(\gamma)$ such that
 $$A(\gamma) U(g)A(\gamma)^{-1} = U(\gamma(g)),\; \gamma
\in Aut({\cal G}).\eq{9}$$ If $A(\gamma)$ is not one of
the
$U(g)$'s we can enlarge the group ${\cal G}$ to include
it. Indeed if
$\gamma$ belongs to a subgroup $\Gamma$ of
$Aut({\cal G})$ then, by virtue of (9), all possible
products of
$A(\gamma)$'s and $U(g)$'s can be written in the form
$U(g)A(\gamma)$ with some choice of $\gamma$ and
$g$. In cases of interest this decomposition will be
unique (i.e.\ the group will be a semi-direct product).
We can then adopt the ``\Schrod picture" and say that
dynamical relations occur between  detectors
$\bra{g}$ and systems that evolve
 according to the dynamical law 
$\ket{0} \to A(\gamma)\ket{0}$ as $\gamma$ follows a
one-parameter subgroup of
$Aut({\cal G})$. Thus amplitudes will have the form
$\bra{g}\gamma\rangle = \bvacp U(g)A(\gamma)\kvac$.
\vskip.1in
Since WH groups are generated by the exponentials of 
linear forms in the bose operators, one sees from the
algebra of bose operators that the exponentials of
{\it quadratic} forms in the bose operators all generate
automorphisms. Indeed all of the linear canonical
transformations are produced in this way, and we are
therefore able to account for all linear dynamical
processes within $Aut({\cal G})$
\vskip.05in 
The following example of such a dynamical
process is of great interest because it illustrates the
mechanism by which entanglement is generated. 
Let us
think of detectors for a pair of laser modes $a,b$ as
a single detector $\bra{\lambda} = \bvac
U^\dagger(\lambda),\; \lambda = (\lambda_a,\lambda_b)$.
The coherence group is ${\cal G} = WH_2$, and among the
canonical automorphisms are those defined by
$$
{\cal D}(\xi) \equiv e^{\xi a^\dagger b^\dagger - \xi^*
a b},\quad \xi
\in {\cal C}.\eq{10}$$
One can show$^{\bf 5}$ that
${\cal D}$ has the normal ordered form:
$$ {\cal D}(\xi) = \beta e^{\zeta a^\dagger b^\dagger}
e^{-
\zeta^* a b}, \quad \zeta = e^{i\arg\xi}\tanh|\xi|,\quad
\beta = \sqrt{1 - |\zeta|^2},$$ so that
$${\cal D}(\xi)\kvac =
 R_\zeta\kvac \equiv \ket{\zeta},\quad R_\zeta = \beta
e^{\zeta a^\dagger b^\dagger}.\eq{11}$$ 
\vskip.1in
The states $\ket{\zeta}$ are two-photon laser states (not
to be confused with two mode laser states).  A simple
manipulation then gives the following amplitude for the
relation between the detector and the reference state as the
two photon laser is ``turned on" by making $\zeta$ non-zero:
$$\bra{\lambda}\zeta\rangle  =
\beta e^{-(|\lambda_a|^2 +|\lambda_b|^2)/2}e^{\zeta
\lambda_a
\lambda_b}.\eq{12}$$ As
$\zeta$ is turned on {\it the presence of the last factor
causes the size of the relation to change in a manner that
depends both on the modulus and phase of $\zeta$.} For fixed
non-zero $\zeta$ the size of the relation does not depend
on $\lambda_a$ and $\lambda_b$ independently --- there is a
correlation which becomes more and more sensitive to the
phase of $\zeta$ as $\zeta$ approaches the boundary of the
unit disc, i.e.\ 
   $\zeta \to
e^{i\phi}$. In this limit 
 the probability tends to
$$p = \beta^2  e^{-|\lambda_a +
e^{i\phi}\lambda_b|^2}.\eq{13}$$
We cannot actually attain the limit because
$\beta$ also tends to zero.
\vskip.05in
The phase of $\zeta$  is influenced by the
space-time locations $x_a,x_b$ associated with the two
laser photons.  If their $4$-momenta are $k_a,k_b$ the
space-time translation automorphism generated by the
$4$-momentum oprator $P = k_a a^\dagger a + k_b b^\dagger
b$ will multiply the bose operators by phases
$e^{ik_a\cdot x_a}$ and
$e^{ik_b\cdot x_b}$ which, as one sees from (11), is
equivalent to multiplying $\zeta$ by the product of the
phases. If the two photons separate 
along any ray in space-time the phase of $\zeta$ will
change by a factor $e^{ik\cdot x}$ where $x$ is the
 separation between them. Thus the correlation
will change with $x$ {\it regardless of
whether the  direction of $x$ is  space-like,
light-like, or time-like.} 
The EPR problem of how to understand 
the correlations when the ray is space-like is now
seen within this manifestly covariant framework to be no
different than how to understand them when the ray is
time-like. 
\vskip.05in
To achieve this understanding let us first remark 
 that  
$\bra{\lambda}$ are coherent states of $WH_2 =
WH_1\otimes WH_1$. While coherent states cannot be added
to produce coherent states, they can always be tensored
to form coherent states. To see this simply note that
the coherence group will be the direct product, and the
representation $U$ will be the tensor product. The
reference state, however, may or may not be the
tensor product of the two reference states. If it is the
tensor product, one will easily check that amplitudes
will factorize, and there will be no correlation. What
happens to produce the correlation in (12) is that for
non-zero $\zeta$ the state
$\ket{\zeta}$ no-longer factorizes.
\vskip.1in
To understand how this non-factorizability can
be be felt between arbitrary space-time points, observe
the remarkable fact that $\ket{\zeta}$ is  a
distorted form of the Fock vacuum. To see why note
that while it is not annihilated by
$a,b$, it is annihlated by the commuting bose operators
$$a(\zeta) = \beta^{-1}(a - \zeta b^\dagger),
\;\; b(\zeta) = \beta^{-1}(b - \zeta
a^\dagger).\eq{14}$$
The vacuum is ``non-local" in the following precise
sense. It is known from a powerful theorem due to
Perelomov$^{\bf 6}$ that the set of $WH$
coherent states $\ket{\lambda}$ with $\lambda$ on
the lattice $n + im$ will be  overcomplete by one, i.e.\
there is exactly one relation of linear dependence. This
can be expressed as as an expansion of the vacuum $\ket{0}$
as a linear combination of all of the others, and in this
exapansion {\it all of the coefficients have the same
modulus}. (That such a state can be normalized is possible
because the coherent states are not mutually orthogonal.)
 Thus the vacuum state is uniformly spread over the
Hilbert space. Distortions in the vacuum therefore express
themselves in a uniform manner everywhere in the space of
coherent states.
\vskip.05in
The distortion becomes greater as
$\zeta$ moves toward the boundary of the unit circle. The
transformations produced by  $D(\zeta)$ are isomorphic
to the $1-1$ Lorentz group with $\zeta$ acting like a
 velocity parameter which reaches lightspeed on
the unit circle. In some sense one can say
that the two photon laser ``boosts" the vacuum, and
the resulting distortion of the reference (analogous to
Fitzgerald contraction) alters the sizes of relations
regardless of the space-time locations of the detection
events. 
\vskip.05in
One may observe that as
$\zeta$ approaches the boundary  the
distorted reference states become sharply distinguishable
from one another in
the following sense: The scalar product is found to be
 $^{\bf 7}$
$$\bra{\zeta^\prime}\zeta\rangle = { {(1 -
|\zeta^\prime|^2)^{1/2}(1 - |\zeta|^2)^{1/2}}\over {1 -
\zeta^{\prime *}\zeta}} \eq{15}$$ which vanishes if 
 either $\zeta$ or $\zeta^\prime$ tend to the boundary, 
unless
$\zeta$ and $\zeta^\prime$ are equal, in which case it
tends to unity. Thus such states with infinitesimally
different phases are sharply distinguished, an indicator
of classical behavior.
 This also means that the relative phase of $\lambda_a$
and
$\lambda_b$ can be sharply measured by the correlation.
Our ability to exploit this, however, is limited by
the vanishing of $\beta$ at the boundary. 
 \vskip.05in
Because ${\cal D}(\xi)$
is an automorphism, the stability subgroup of
$\ket{\zeta}$ is isomorphic to that of the Fock vacuum.
We are naturally led to consider what sort of
entanglement will result if the stability subgroup of the
reference state for a direct product is non-isomorphic
to that of the factors. Bell states
such as the Bohm-Aharonov singlet arise in this way,
and indeed states of this kind are associated with 
coherent states of every compact
${\cal G}$ in the following way:$^{\bf 8}$  
\vskip.05in Observe first that the unit operator can be
written
$$ I = \int_{\cal S} \ket{g}\bra{g} d\mu,
\eq{16}$$ where
$\bra{g}$ and $\ket{g}$ are duals of one another). This
may be deduced from Schur's Lemma  noting that it
commutes with every element
$U(g)$ of an irreducible representation. This also
establishes the (over) completeness of the coherent
states noted earlier. Now suppose we have an
anti-unitary map
$$\ket{g} \to \ket{g^*} \Longrightarrow
\bra{g^*}h^*\rangle =
\bra{h}g\rangle,\eq{17}$$ and define the tensor product
state
$$\doubleket{{\cal B}} \equiv \int_{\cal F}d\mu
\ket{g}\otimes\ket{g^*}.\eq{18}$$ The stability subgroup
for this state {\it is not the direct product of those
of the constituents}. Now the state is  invariant under
all transformations of the form
$(h,h^*)$ (with the same
$h$ in each component). Consider the effect of a
change in reference state from $\kvac = 
\ket{0_1}\otimes\ket{0_2}$ to
$\doubleket{{\cal B}}$ on the relation between the
reference state and a  detector state
$\bra{g_1,g_2^*} = \bra{g_1}\otimes
\bra{g_2^*}.$ We see that 
$$\bra{g_1,g_2^*}0\rangle =
\bra{g_1}0_1 \rangle
\bra{g_2^*}0_2\rangle\Longrightarrow
\langle g_1,g_2^*\doubleket{{\cal B}} =
\int_{\cal F}d\mu
\bra{g_1}g\rangle\bra{g_2^*}g^*\rangle =
\bra{g_1}g_2\rangle,\eq{19}$$ thus producing {\it the
same correlation we would have if the two constituents
of the detector belonged to the same Hilbert space.} In
particular the probability is unity for the relation to
hold when
$g_1 = g_2$ i.e.\ the state $\doubleket{{\cal B}}$
always passes a pair of detectors in conjugate states.
\vskip.05in Since 
$$\langle\langle{\cal B}||{\cal B}\rangle\rangle =
\int_{\cal F}d\mu,\eq{20}$$
\noindent we see that the generalized Bell state has a
finite norm if and only if ${\cal F}$ is compact. We
shall therefore be able to construct generalized Bell
states in this way for compact groups only. Note that the
construction shows clearly why anti-unitary
transformations are always present in the construction
of Bell states. 
\vskip.05in The
generalized Bell state computed from (18) for
${\cal F}= SU_2/U_1$ is found to be the Bohm-Aharonov
singlet. The anti-unitary transformation $*$ is time
reversal. 
Although one cannot directly apply (18) to the $WH_2$ group
because it is not compact, it is noteworthy, that the states
$\ket{\zeta}$ can be produced by inserting a gaussian
convergence factor parametrized by $\zeta$ into (18). 
\vskip.05in
I have now demonstrated that the CS model can account for
 the phenomena which
have been taken to justify the superposition principle.
 In the CS model the classical proposition
that a dynamical variable lies in a certain phase space
neighborhood is replaced by the proposition that a
relation
$g$ in a certain neighborhood of ${\cal S}$ ``holds". We
have seen that the CS model admits a relational hidden
variable, i.e.\ a Kolmogorov probability space that
accounts for the statistical distribution of outcomes by
a simple kind of noise in the space of relations.
\vskip.05in In the CS model there is no sharp separation
between system and detector.  A relation $g_1^{-1}g_2$
does not have a unique factorization, and the complete
symmetry between system and detector is seen in the
invariance of probabilities under complex conjugation
which exchanges their roles in amplitudes. 
\vskip.05in One may contrast these features of the CS
model with the orthodox (Born-Dirac-von Neumann) model
in which ``properties" are identified with lattices of
subspaces in Hilbert space$^{\bf 9}$. The no-go theorems,
e.g.\ Kochen-Specker $^{\bf 10}$) tell us that we cannot
assign values (hidden variables) to
account for the observed distribution of outcomes of
measurements made to determine such properties.
The familiar paradigm for the interaction of a system
with a measuring device produces an
entangled state  which must quickly ``collapse"
if a determinate value for the property is to be obtained.
The failure of the orthodox interpretation to account
for the interaction without a non-linear modification to
produce collapse is called the {\it measurement problem}.
In the CS model the notion of ``property" is  replaced
by the $g$-value of a relation, and we then have a hidden
variable account of what may happen in a test of the
relation to explain the statistical distribution of
outcomes.
 \vskip.05in The CS model recognizes entangled states as
distorted vacuum states which can evolve in a manner that
is completely consistent with special relativity but
  nonetheless induce correlations between space-like
events. The fact that a distortion of the vacuum  effects 
all event pairs regardless of
whether they are time-like, light-like, or space-like to
one another is explained by the uniform non-locality of the
vacuum.
\vskip.05in 
The basic question we must answer before claiming that we
have resolved the measurement and EPR problems
 is whether a coherence
group exists for all quantum phenomena. The similarity of
the double cosets structure to vacuum expectation values and
the field theoretic form of the general tensor products of
WH groups suggests that the way to find the coherence group
for any system is to second quantize. Thus it is plausible
to suppose that the way to guarantee that there is no
measurement problem is to frame the theory as one of
relations between regulated quantum fields.

\vskip.2in

\centerline{{\bf References}} 1. R. Glauber, {\bf
131},2766,(1963)

2. A. Perelomov, ``Generalized Coherent States and their
Applications", 

\hskip.2in  Springer-Verlag (1986)

3. M. Hall, Jr., ``The Theory of Groups"  p. 14,
The Macmillan Co.

\hskip.2in (1959).

4. J.S. Bell, Physics {\bf 1 }, 195 (1964)

5. A. Perelomov, op. cit. p. 73 

6. A. Perelomov, op. cit. p. 25

7. A. Perelomov, op. cit. p. 74 

8. D. Fivel, UMBC Workshop on Fundamental Problems in
Quantum 

\hskip.2inTheory. ``The Relationship Between Coherence
and Entanglement" 

\hskip.2in (1999).

9. J. Bub, ``Interpreting the Quantum World",
p. 24 and p. 31 et. seq. 

\hskip.2in Cambridge University  Press, (1997)

10. S. Kochen an E.P. Specker, J. Math. Mech. {\bf 17},
59, (1967)
\end